\documentclass[twocolumn, showpacs,amsmath,amssymb]{revtex4}
\usepackage{epsfig}
\usepackage{graphicx}
\usepackage{epstopdf}
\usepackage{bm}

\usepackage{color}

\begin{document}
\title{Quantum driving protocols for a two level system: from generalized Landau-Zener sweeps to superadiabatic control}
\author{N. Malossi$^{1}$, M. G. Bason$^2$\footnote{present address: Department of Physics and Astronomy, Aarhus University Aarhus, Denmark }, M. Viteau$^2$\footnote{present address: Orsay Physics, 95 av. des monts AurŽliens, ZAC Saint Charles, 13710 Fuveau - France}, E. Arimondo$^{1,2,3}$,  R. Mannella$^1$, O. Morsch$^2$, and D. Ciampini$^{1,2,3}$}
\affiliation{$^1$Dipartimento di Fisica `E. Fermi', Universit\`a di Pisa, Largo Pontecorvo 3, 56127 Pisa, Italy\\ $^2$INO-CNR, Largo Pontecorvo 3, 56127 Pisa, Italy\\ $^3$CNISM UdR, Dipartimento di Fisica `E. Fermi', Universit\`a di Pisa, Largo Pontecorvo 3, 56127 Pisa, Italy \\}

\begin{abstract}
We present experimental results on the preparation of a desired quantum state in a two-level system with the maximum possible fidelity using driving protocols ranging from generalizations of the linear Landau-Zener protocol to transitionless driving protocols that ensure perfect following of the instantaneous adiabatic ground state. We also study the minimum time needed to achieve a target fidelity and explore and compare the robustness of some of the protocols against parameter variations simulating a possible experimental uncertainty. In our experiments, we realize a two-level model system using Bose-Einstein condensates inside optical lattices, but the results of our investigation should hold for any quantum system that can be approximated by a two-level system.
\end{abstract}

\maketitle
\section{Introduction}
The "simplest non-simple quantum problem"  is the evolution of a two-level quantum system with a time-dependent Hamiltonian~\cite{Berry_95a}. A paradigmatic example of such a system is the Landau-Zener problem, in which two levels whose energy depends on time through some parameter  experience an avoided crossing during the temporal evolution. If the time variation of the energy levels is linear, the resulting evolution gives rise to the well-known phenomenon of Landau-Zener (LZ) tunneling~\cite{Landau1932,Zener1932,NoteMajorana},  a scenario lying at the heart of a host of quantum phenomena. The LZ tunneling probability is often used to estimate the degree of adiabaticity achievable when a two-level system is forced through an avoided crossing of its energy levels. In modern terminology, this can be described as a problem in quantum control: if the goal is to keep the system as close as possible to the lowest energy level before and after the crossing, one has to choose a time dependence that minimizes the probability of LZ tunneling. In other words, one aims to maximize the fidelity of the control protocol, defined as the overlap between the desired and the actual final state. For the standard LZ problem with a linear time-dependence of the energy, the goal of perfect preparation of the final state can only be achieved for infinitely slow sweep speeds or for infinitely large coupling strengths, both of which are impractical (and, indeed, unphysical).\\
\indent This observation leads naturally to the following question: can the degree of fidelity in a LZ type quantum control protocol be improved by using non-linear sweeps in some control parameter $\Gamma$? If the off-diagonal element $\omega$ of the Hamiltonian can be also varied during the protocol, what is the optimum combination of the Hamiltonian parameters that maximizes the fidelity of the final state? Finally, is it possible to choose the parameters of the two-level system in order to speed up the transfer from the initial state to the final one? All these questions represent key issues within the present effort to control quantum phenomena in different areas from atomic physics to solid-state physics and molecular physics~\cite{WamsleyRabitz2003,Rice2000}.\\
\indent The quest for optimal control of the two-level systems has a long history, and protocols denoted as Rosen-Zener, Allen-Eberly or Demkov-Kunike have been studied for many years~\cite{GarrawaySuominen1995}. In magnetic resonance, composite pulses were introduced to create fast and robust  transfer protocols~\cite{Levitt1996}. Within the framework of adiabatic quantum computation, Roland and Cerf~\cite{RolandCerf2002} introduced an adiabatic protocol that is much faster than the standard LZ protocol.  Protocols with nonlinear time dependence of the $\Gamma$ parameter, the so-called generalized LZ sweeps, have been investigated because, in contrast to the LZ problem with a linear sweep, they alllow a complete transition to the target state with a finite sweep rate~\cite{GaraninEL2002,GaraninPRB2002}. An alternative approach is  to use shortcut-to-adiabaticity protocols leading to the same target state as the LZ approach, but in a shorter time.  Several methods to find shortcuts to adiabaticity have been proposed for two-level systems, such as reverse engineering using the Lewis-Riesenfeld invariants~\cite{Muga2009,Muga2012}. Finally, the super-adiabatic (also known as counter-adiabatic or transitionless) protocols for a given time-varying Hamiltonian construct an auxiliary Hamiltonian  that exactly cancels the non-adiabatic part of the original Hamiltonian and thus ensures perfect adiabatic following~\cite{Lim1991,Berry2009,Demirplak2003,Demirplak2008}.\\
\indent A related issue is the desire to (accurately) realize  these quantum manipulations in the shortest possible time.  There is, however, a fundamental limit towards this goal which is  set by quantum mechanics. There is a maximum speed, imposed by the  Schr\"odinger evolution, at which a quantum system can evolve
~\cite{levitin,bhattacharyya,peres,Giovannetti2003,caneva}.  If a quantum system can be controlled by a  set of external parameters, the speed limit can be attained only for a suitably tailored time-dependence of these control parameters.\\
\indent All the above questions have so far received limited experimental attention.   The Roland-Cerf protocol was experimentally implemented on different quantum computation algorithms by nuclear magnetic resonance techniques in~\cite{Mitra2005}.   The control of nonadiabatic transitions in an adiabatic transformation was performed  in the experiment of ref. ~\cite{Lu2007}. A nonadiabatic transformation leading to a nearly perfect preparation of the target state was realized in an experiment  on a Bose-Einstein condensate within an optical lattice in~\cite{Mellish2003}. Our recent work with Bose-Einstein condensates compared the efficiency reached in the LZ, Roland-Cerf and superadiabatic protocols~\cite{Bason2012}. The shortcuts to adiabaticity based on the the Lewis-Riesenfeld invariants for the case of Hamiltonian with a dependence also on the spatial coordinates  were investigated in experiments on trapped ultracold gases~\cite{Schaff2011}.\\
\indent The quantum speed limit had not been explicitly tested before our investigation in ref.~\cite{Bason2012}. In nuclear magnetic resonance or atomic physics, it is typically assumed that the most rapid transfer is produced by a $\pi$ Rabi pulse. In fact, for a two-level Hamiltonian with constant parameters, the $\pi$ pulse corresponds to the quantum speed limit. However, for more complicated protocols such as those investigated here, the deviation from the quantum speed limit is an important issue.\\
\indent The present work reports an experimental investigation of different protocols regarding their fidelity and the deviation from the quantum speed limit. Section II briefly introduces the quantum driving of our two level paradigmatic quantum system and the also the quantum speed limit. Starting from the linear LZ as a reference protocol investigated in~\cite{Zenesini2009,Tayebirad2010}, we explored generalized LZ protocols where $\omega$  is fixed and $\Gamma$ is scanned non linearly (Section III). Section IV discusses the Roland and Cerf protocol. Section V introduces the idea of superadiabatic protocols and reports time-dependent measurements for both the diabatic and adiabatic bases. Section VI investigates non-optimized Roland and Cerf protocols and superadiabatic protocols, for a deviation of control parameters from optimal values.  Section VII discusses the best use of resources, which are always limited experimentally, in terms of the minimum required to reach a certain target fidelity. Section VIII presents the conclusions of our experimental and theoretical investigations.

\section{The system and the target}
\subsection{Hamiltonian}
As described in the Appendix and shown schematically in Fig. \ref{bandstructure}, our system is realized with ultracold atoms forming a BoseÐ Einstein condensate (BEC), in an accelerated optical lattice.  Under appropriate conditions the BEC can be described as a two-level system defined by the two lowest Bloch bands:
\begin{equation}
	{\cal H} =  \hbar \Gamma (t) \sigma_z + \hbar \omega(t) \sigma_x,
\label{Hamiltonian}
\end{equation}
where $\Gamma$ is the energy between the diabatic components of the Hamiltonian and $\omega$ is the intensity of the coupling between the diabatic components, causing the splitting between the bands. The above generalized time-dependent LZ Hamiltonian has instantaneous eigenvectors $|\psi_\mathrm{g,e}(t)\rangle$, its \emph{adiabatic} eigenstates. The \emph{diabatic} states $|0\rangle,|1\rangle$ are the eigenvectors of the Hamiltonian with $\omega(t)=0$.\\
\indent The case  of constant $\omega$ and  $\Gamma(t)$ depending linearly on $t$ corresponds to the standard LZ Hamiltonian~\cite{Landau1932,Zener1932,Majorana1932} with an  avoided crossing with a minimum energy gap of $2\hbar \omega$ at  $\Gamma(t)=0$.  For our system of a BEC in an optical lattice, the lattice depth controls the tunneling barrier and hence the gap, while an acceleration of the lattice controls the time dependence of $\Gamma$. \\
\indent  Having prepared the system in the ground state of ${\cal H}$ for a given choice of the initial parameters $\omega_i$ and $\Gamma_i$ we would like to reach, after an evolution of duration $T$, the ground state of ${\cal H}$ for the final parameters $\omega_f$ and $\Gamma_f$.  The initial and final values of the parameters are chosen to be on opposite sides of the anticrossing.  The different protocols examined in the following correspond to different time dependencies $\Gamma(t)$ and $\omega(t)$.\\
\indent  For simplicity, $\Gamma$, $\omega$ and $T$ are dimensionless, using the natural units of energy and time in our Rubidium system, see Appendix A. In addition, for a given value of $T$,  we introduce the rescaled time $\tau=t/T$, $\tau \in  [0,1]$, and we assume that the sweep ranges from  $-\Gamma_0$ to $\Gamma_0$, with $\Gamma_0=2$, and is point-symmetric about $\tau=0.5$. \\

\begin{figure}[t]
\centerline{\includegraphics[angle=0., width = 0.9\linewidth] {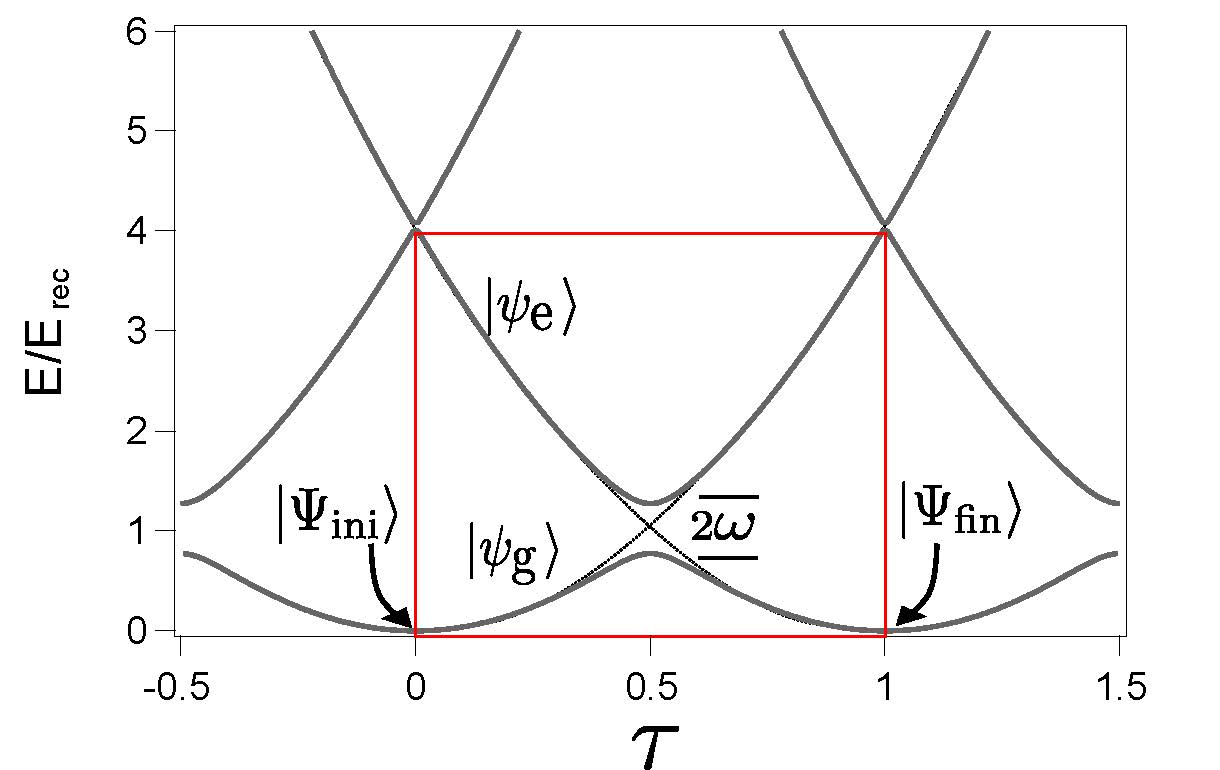}}
\caption{Experimental realization of an effective two level system exploiting the band structure of a Bose-Einstein condensate in an optical lattice. The avoided crossing is located at the edge of the Brillouin zone and the two adiabatic states $|\psi_\mathrm{g,e}(\tau)\rangle$ are the lowest energy bands. The initial and final states of the time evolution $|\Psi_\mathrm{ini,fin}\rangle$ are also indicated.}
\label{bandstructure}
\end{figure}

\subsection{Fidelity}
 We characterize the protocol efficiency  for the system  going from the initial state $|\Psi_\mathrm{ini}(\tau=0) \rangle=|\psi_\mathrm{g}(\tau=0)\rangle$ to the final state $|\Psi_\mathrm{fin}(\tau=1)\rangle$ through the following final fidelity:
\begin{equation}
\mathcal{F}_\mathrm{fin}= |\langle \psi_\mathrm{g}(\tau=1)| \Psi_\mathrm{fin}(\tau=1)\rangle |^2.
\end{equation}
The evolution of the wavefunction along the adiabatic trajectory at an intermediate time $\tau$ for given $T$ may be tested by extending the above definition  to a time dependent fidelity $\mathcal{F}(\tau)$. In the experiment the wavefunction at time $\tau$ is determined by a projective measurement on the adiabatic or diabatic bases of the Hamiltonian of Eq.~\eqref{Hamiltonian}. Notice that for the case of the diabatic basis, the projective measurement of the wavefuction $|\psi_g\rangle$ results in the following  probability:
\begin{equation}
{\mathcal{P}}^\mathrm{diab}(\tau)= |\langle 1| \psi_\mathrm{g}(\tau)\rangle|^2,
\end{equation}
 where we denote by $|0\rangle$ the state that at $\tau=0$ has the lowest energy. For the superadiatic protocols  the wavefunction follows precisely the instantaneous adiabatic eigenvalues of the Hamiltonian of Eq.~\eqref{Hamiltonian}, so that the above probability becomes
 \begin{equation}
{\mathcal{P}}^\mathrm{diab}(\tau)= \frac{\omega^2(\tau)/2}{\Gamma^2(\tau)+\omega^2(\tau)+\Gamma(\tau)\sqrt{\Gamma^2(\tau)+\omega^2(\tau)}}.
\label{ProbDiab}
\end{equation}

\subsection{Quantum speed limit}
The quantum speed limit (QSL) protocol  allows us to take the system from the initial  state to the final state, with final fidelity  $\mathcal{F}_\mathrm{fin}= 1$, in the  shortest possible time $T$. This lower bound is rooted in the Heisenberg uncertainty principle~\cite{levitin,bhattacharyya,peres,Giovannetti2003,caneva}. For the case of constant $\omega$,  we found that the protocol minimizing  $T$ is~\cite{Bason2012}
 \begin{eqnarray} \label{ggg}
\Gamma(\tau) = \left\{
\begin{array}{ll}
-\Gamma_0   & \mbox{for $t=0$} \\
\Gamma_M  & \mbox{for $t\in[0,t_0]$} \\
0 & \mbox{for $t\in[t_0,T-t_0]$} \\
-\Gamma_M   & \mbox{for $t\in[T-t_0,1]$} \\
+\Gamma_0   & \mbox{for $t=T$}\;, \\
\end{array}\right.
\label{qsl}
\end{eqnarray}
where  $\Gamma_M$ and $t_0$ are, respectively, asymptotically large and small quantities which satisfy the condition $\Gamma_M t_0 = \pi/4$.\\
\indent  This `composite pulse' protocol, in close analogy to composite pulses in NMR~\cite{Levitt1996}, represents half a Rabi oscillation with frequency $\omega$ and $\Gamma$=0, preceded and followed by two short pulses (in theory delta-functions) with a pulse area of $\pi/4$.
 The transfer time $T$ associated with the $|\Psi_\mathrm{ini}\rangle$ to $|\Psi_\mathrm{fin}\rangle$ transfer through the above protocol is
\begin{eqnarray}
T = 2t_0+ \frac{\arccos{|\langle \Psi_\mathrm{fin}|\Psi_\mathrm{ini}\rangle|}}{ \omega}\;.
\label{qsloptimal}
\end{eqnarray}
\\
\indent  A Rabi rotation of an angle $\pi$ corresponds to a time $T_\pi=\pi/\omega$, and in fact in the limit of $t_0 \to 0$, the above time coincides with $T_\pi$. For $\omega$ values up to  1, the $T_{\pi}$ dependence on $\omega$  matches the lower continuous line of Fig.~\ref{LZoptimized}. At larger $\omega$ values, $|\psi_g(\tau=0)\rangle$ differs from $|0\rangle$, and the initial tilting angle $\theta=\omega/\Gamma_0$ of the Bloch  vector leads to  a required Rabi rotation smaller than $\pi$.

\section{Generalized LZ sweeps}
We begin our investigation of the dynamics of generic two-level systems starting from the simplest possible generalizations of the LZ scenario. In this Section we keep the coupling $\omega$ constant and study deviations of $\Gamma(t)$ from the linear dependence assumed in the original LZ problem. We introduce nonlinear sweeps in two ways: by making the functional form of $\Gamma(\tau)$ follow some power law in $\tau$ and by adding a small nonlinear contribution to the original linear sweep.

\begin{figure}[t]
\centerline{\includegraphics[angle=0., width = 1.1\linewidth] {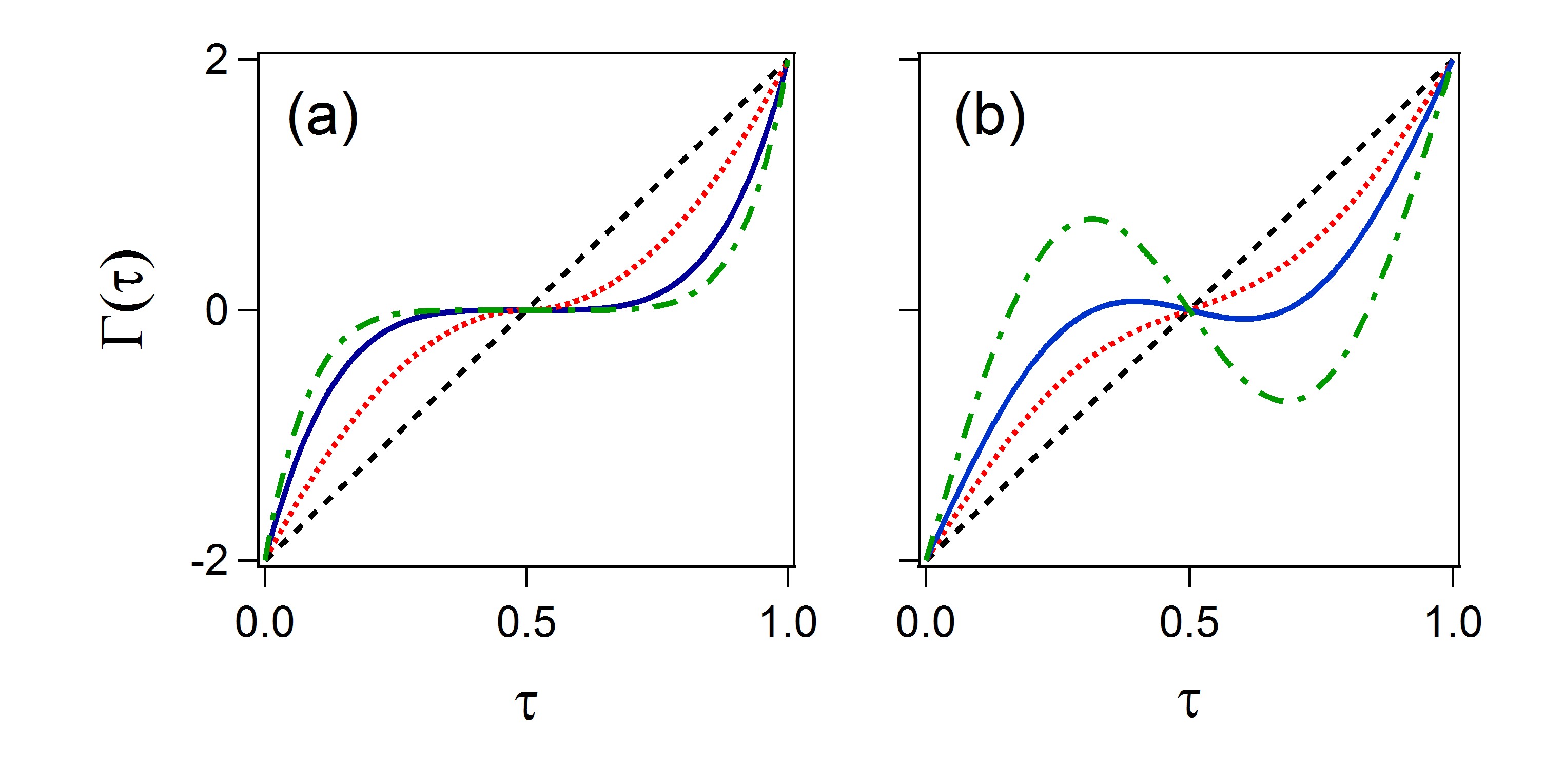}}
\caption{Time dependence of the control parameters for different protocols. In (a) $\Gamma(\tau)$ for power law sweeps: red dotted line $\alpha= 2$, blue continuous line $\alpha= 3$, green dot-dashed line $\alpha= 4$.  In (b)  $\Gamma(\tau)$ for linear+sin sweeps: red dotted line $\delta= 0.1$, blue continuous line $\delta= 0.2$, green dot-dashed line $\delta= 0.4$. The black dashed line corresponds to the linear LZ sweep.}
\label{GammaDependencies}
\end{figure}
\subsection{Power laws}
\begin{figure}[t]
\centerline{\includegraphics[angle=0., width = 0.9\linewidth] {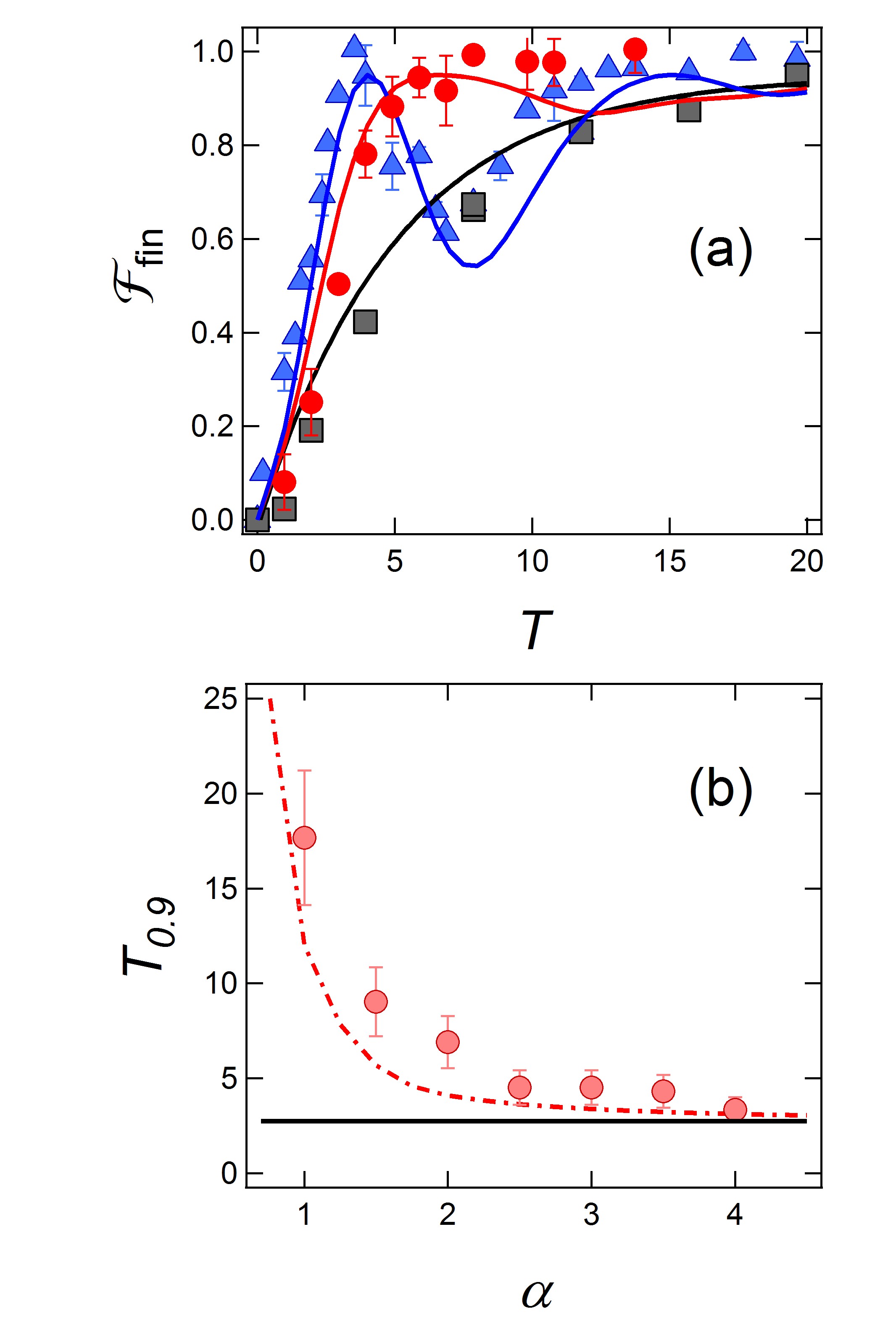}}
\caption{(a) Final fidelity as a function of the  duration $T$ of the power law sweep. Blue triangles $\alpha=4$, red circles $\alpha=2$, black squares  for linear LZ with $\alpha=1$.  The continuous lines (blue, red and black) represent numerical simulations for the corresponding power law sweeps.  (b) Minimum time $T_{0.9}$ to achieve $\mathcal{F}_\mathrm{fin}= 0.9$ for the power law sweeps. The experimental points (solid circles) are compared to a numerical simulation (dot-dashed red line). The black line shows the minimum time necessary to achieve $\mathcal{F}_\mathrm{fin}= 1$ according to the quantum speed limit. All data with $\omega=0.5$}
\label{powerdata}
\end{figure}

The first scheme leads to sweeps of the following form satisfying the boundary conditions $\Gamma(0)=-2$ and $\Gamma(1)=2$:
\begin{eqnarray} \label{power}
\Gamma(\tau) = \left\{
\begin{array}{ll}
-2^{\alpha +1}($1/2$-\tau)^{\alpha}   & \mbox{for $0 \leq \tau \leq 1/2$} \\
2^{\alpha +1}(\tau - $1/2$)^{\alpha}   & \mbox{for $1/2 \leq \tau \leq 1$}\;, \\
\end{array}\right.
\end{eqnarray}
The case $\alpha= 1$ corresponds to the linear LZ sweep and different power laws lead to the time dependencies depicted in Fig.~\ref{GammaDependencies}(a). For $\alpha >1$ the bandgap at the edge of the Brillouin zone is crossed once with zero speed, and for increasing values of $\alpha$ the width of the low speed region increases, i.e. $|\Gamma(\tau)|\leq C$, for $|\tau-1/2| \leq (C/2)^{1/\alpha}$. For example, if $\alpha=4$ and $C= 10^{-3}$, the width of the low speed region is 15$\%$ of the total sweep duration.\\
\indent  Power law sweeps have been studied theoretically by Garanin and Shilling ~\cite{GaraninPRB2002,GaraninEL2002}. They calculated the corrections to the LZ transition probability due to the nonlinearity, finding that it has an   oscillatory behavior if the sweeps exhibits a significant retardation in the vicinity of the resonance, i.e. $\dot{\Gamma}(\tau)$ is close to zero around $\tau=0.5$, in agreement with our experimental observations.\\
\indent We have examined experimentally and numerically the dynamics of a two level system for various values of $\alpha$. In Fig.~\ref{powerdata}(a) the fidelity of the final state for $\alpha=1,2,4$ is plotted as a function of the total duration $T$ of the sweep, keeping  $\omega =0.5$ fixed. Numerical simulations of the power laws sweeps shown in Fig.~\ref{powerdata}(a) agree well with the experimental data. Sweeps with larger values of $\alpha$ reach a given final fidelity in a shorter time. Fixing a threshold value at $\mathcal{F}_\mathrm{fin}= 0.9$, the required sweep duration $T_{0.9}$ strongly depends on $\alpha$, as shown in Fig.~\ref{powerdata}(b). For large values of $\alpha$, $T_{0.9}$ approaches the minimum time given by the quantum speed limit for the parameters of our system ($T_{QSL}=2.75)$. In that regime, the protocol approaches the form of the composite pulse protocol of Eq.~\ref{qsl}.

\begin{figure}[h!]
\centerline{\includegraphics[angle=0., width = 1.1\linewidth] {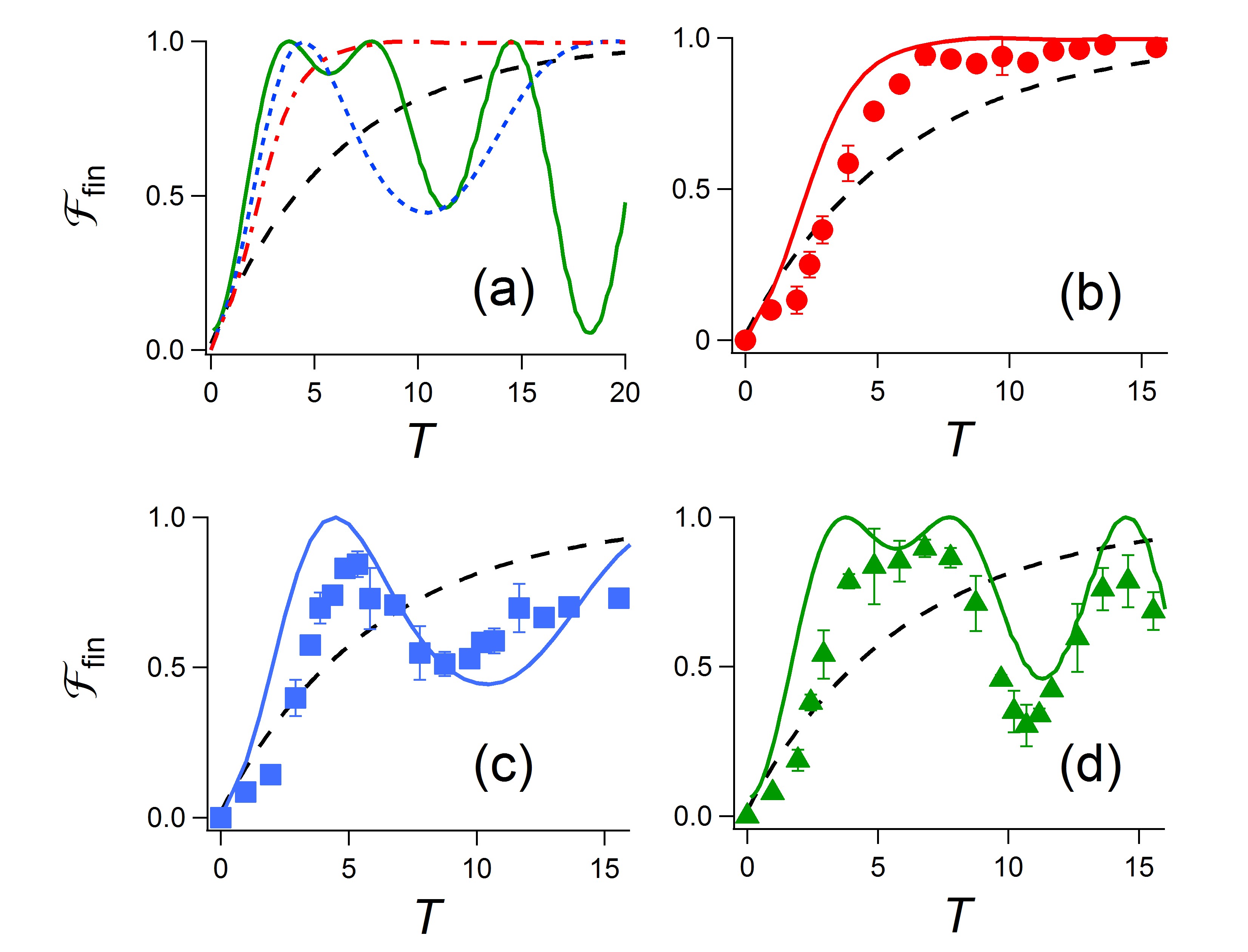}}
\caption{  In (a) numerical simulations for the final fidelity as a function of the  duration $T$ of the linear plus sin sweep, for different values of the  parameter $\delta$ ($\delta = 0.1$ red dash-dotted line, $\delta = 0.2$ blue dashed line, $\delta = 0.4$ green continuous line). For increasing  values of $\delta$ the final state is reached at an earlier time. In (b-d) experimental results are compared to the simulations. In (b) $\delta = 0.1$, in (c) $\delta = 0.2$, in (d) $\delta = 0.4$. Typical error bars are reported. The dashed black line shows the theoretical result for the linear LZ sweep. All data with $\omega=0.5$.}
\label{LinearSin}
\end{figure}

\subsection{Linear+sin}

Nonlinear sweeps can also be created by adding a small sinusoidal contribution to the original linear sweep:
\begin{eqnarray}
\Gamma(\tau) = 4[\tau +\delta\sin(2\pi\tau)]-2\;.
\end{eqnarray}
which satisfies the requirements $\Gamma(0)=-\Gamma(1)=2$ and $\Gamma(0.5)=0$, as shown in Fig.~\ref{GammaDependencies}(b). Depending on the  value of $\delta$ the eigenenergies of the Hamiltonian of Eq.~\eqref{Hamiltonian} are characterized by either one or three avoided crossings during the temporal evolution.  In our experiment the bandgap is crossed only once for $\delta=0.1$ and twice for $\delta=0.2$ and $0.4$. The speed of the crossing at $\tau = 0.5$ depends on $\delta$ as $\dot{\Gamma}(0.5)= 4-2 \pi \delta$. Our measurements and the numerical simulation of Fig.~\ref{LinearSin}(a) show that the fidelity of the final state is not monotonic as a function of the duration $T$ of the sweep. In addition, for the largest $\delta$ the fidelity experiences an oscillatory behavior similar to that of power law sweeps. Those oscillations allow us to reach the final state more rapidly than the linear LZ sweep.

\section{Roland-Cerf  protocol}

\begin{figure}[t]
\centerline{\includegraphics[angle=0., width = 0.8\linewidth] {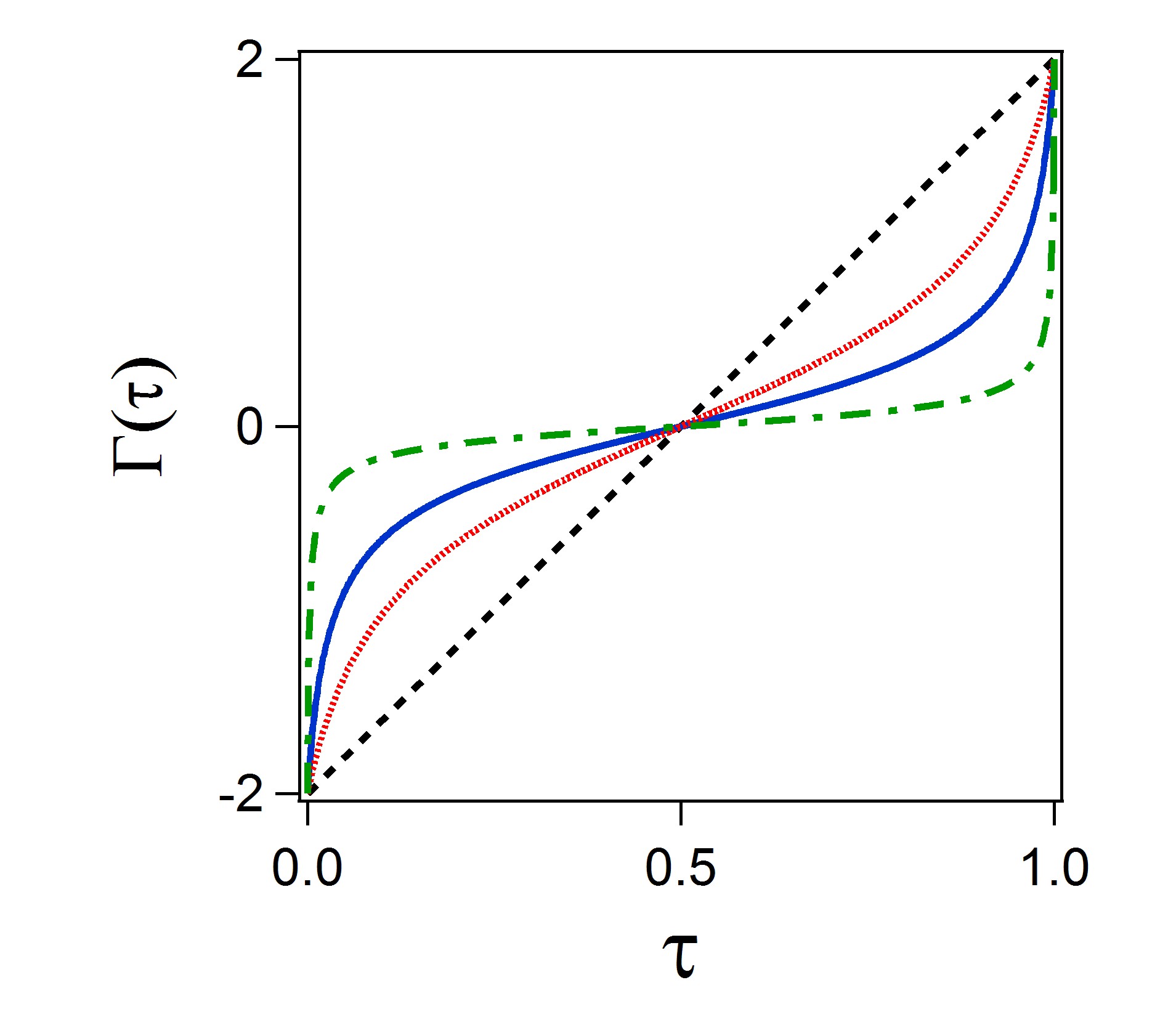}}
\caption{Time dependence of the control parameters for Roland-Cerf protocol: blue $\eta^2= (4+\omega^2)^{-1}$ corresponding to the optimized one, red $\eta^2= 0.85(4+\omega^2)^{-1}$, dashed-dotted green $\eta^2=1.05825(4+\omega^2)^{-1} \approx \sqrt{0.25}$.  The black dashed line corresponds to the linear LZ sweep.}
\label{GammaDependencies3}
\end{figure}
The locally adiabatic protocol proposed by Roland and Cerf~\cite{RolandCerf2002} in the context of adiabatic quantum computation poses stricter constraints on the time evolution of the system. We summarize the analysis in~\cite{RolandCerf2002} adapting it to the case of the Hamiltonian ${\cal H}$ of  Eq.~\ref{Hamiltonian} under the assumption that the coupling $\omega$ is kept constant. At any instant of the evolution the fidelity of the state with the instantaneous ground state is required to be
\begin{equation}
\mathcal{F}(\tau)=  | \langle \psi(\tau) | \psi_\mathrm{g}{(\tau)} \rangle|^2=1 - \epsilon^2,
\end{equation}
It follows~\cite{Bason2012} that the total time of the protocol is
\begin{equation}
T_{F} =  \frac{1}{\epsilon \omega} \; \frac{1}{\sqrt{4 + \omega^2}} \;,
\label{RolandCerftime}
\end{equation}
for a time-dependence of $\Gamma(\tau)$ of the form
\begin{equation}
\Gamma(\tau) = \frac{ 4 \epsilon \omega^2 T_{\cal F} (\tau - 1/2)}{ \sqrt{ 1 - 16 \epsilon^2 \omega^2 T_{\cal F}^2 (\tau - 1/2)^2} }.
\label{RolandCerfGamma}
\end{equation}
\indent We have implemented the above protocol and verified the Eq.~\eqref{RolandCerftime} for the $T_{\cal F}$ dependence on $\omega$~\cite{Bason2012}. However, the realization of the Roland and Cerf (RC) protocol requires a precise knowledge of the parameters of the underlying physical system, since the duration of the sweep $T_{F}$ has to be matched to the coupling constant $\omega$.\\
Moreover, we found that, for a given $\omega$, the RC protocol reaches the target fidelity $\mathcal{F}=0.9$ within a time only twice that of the time-optimal composite pulse protocol.
\\

\section{Superadiabatic protocols}

A further generalization of the LZ scenario is realized by allowing the possibility of a time-varying $\omega(t)$. We have analysed those protocols that ensure a perfect following of the adiabatic ground state $|\psi_\mathrm{g}(\tau)\rangle$ for all $\tau$, called `superadiabatic' or transitionless with reference to Berry's work \cite{Lim1991,Berry2009} and counter-adiadiabatic with reference to the work of Demirplak and Rice~\cite{Demirplak2003,Demirplak2008}. Such protocols  enable fully adiabatic preparation of a desired quantum state of a system in a finite time. The underlying idea is that, given a Hamiltonian $\cal H$, it is always possible to construct an additional control Hamiltonian $\cal H_{\textrm{c}}$ such that the under the action of $\cal H + \cal H_{\textrm{c}}$ the system will remain perfectly in the ground state of $\cal H$ at all times. In principle, $\cal H_{\textrm{c}}$ should be experimentally implemented through an extra control field. In our previous work \cite{Bason2012}, we found that it is possible to eliminate the need for an additional field by suitably transforming the time dependence of the initial Hamiltonian $\Gamma \rightarrow \Gamma^\prime$ and $\omega \rightarrow \omega^\prime$.\\
\indent Two superadiabatic protocols are considered here. The superadiabatic linear protocol, corresponding to the transformation of the linear LZ protocol with $\Gamma(\tau)=4\left(\tau -1/2\right)$, is given by
\begin{eqnarray}
\Gamma^\prime(\tau)&=\Gamma(\tau)-\frac{4(\tau-\frac{1}{2})}{T^2\left[(\tau-\frac{1}{2})^2+\frac{1}{2}\omega^2\right]^2+1}, \nonumber \\
\omega^\prime(\tau)&=\omega\sqrt{1+\frac{1}{T^2(8(\tau-\frac{1}{2})^2+\frac{1}{2}\omega^2)^2}}.
\label{superlin}
\end{eqnarray}
The superadiabatic tangent protocol, corresponding to the transformation of the tangent protocol for which $\Gamma(\tau)$ is not modified by the transformation, is given by
\begin{eqnarray}
\Gamma^\prime(\tau)&=&\Gamma(\tau)=\omega\tan\left(2\left(\tau-\frac{1}{2}\right)\arctan\left(\frac{2}{\omega}\right)\right),\nonumber \\
\omega^\prime&=&\omega \sqrt{1+\frac{\arctan(\frac{2}{\omega})^2}{(T\omega)^2}}.
\label{supertan}
\end{eqnarray}
For both of these protocols the discontinuities in the derivative of $\Gamma$ at $\tau$=0 and $\tau$=1 lead to  delta-functions in $\Gamma^\prime$ at the beginning and at the end of the time evolution, which can be realized in practice using large but finite corrections $\Delta \Gamma_{M}$ for a short duration $\Delta \tau$ such that

\begin{equation}
\Delta \tau \Delta \Gamma_{M} =  \mp \frac{1}{2}\arctan\left(\frac{\Gamma \dot{\omega}- \omega \dot{\Gamma}}{2 \omega (\Gamma^{2}+\omega^{2})}\right)
\end{equation}
where the - and + signs refer to the correction at the beginning and at the end of the protocol, respectively. These superadiabatic protocols are shown schematically in Fig.~\ref{GammaDependencies}(d).

\begin{figure}[t]
\centerline{\includegraphics[angle=0., width = 0.8\linewidth] {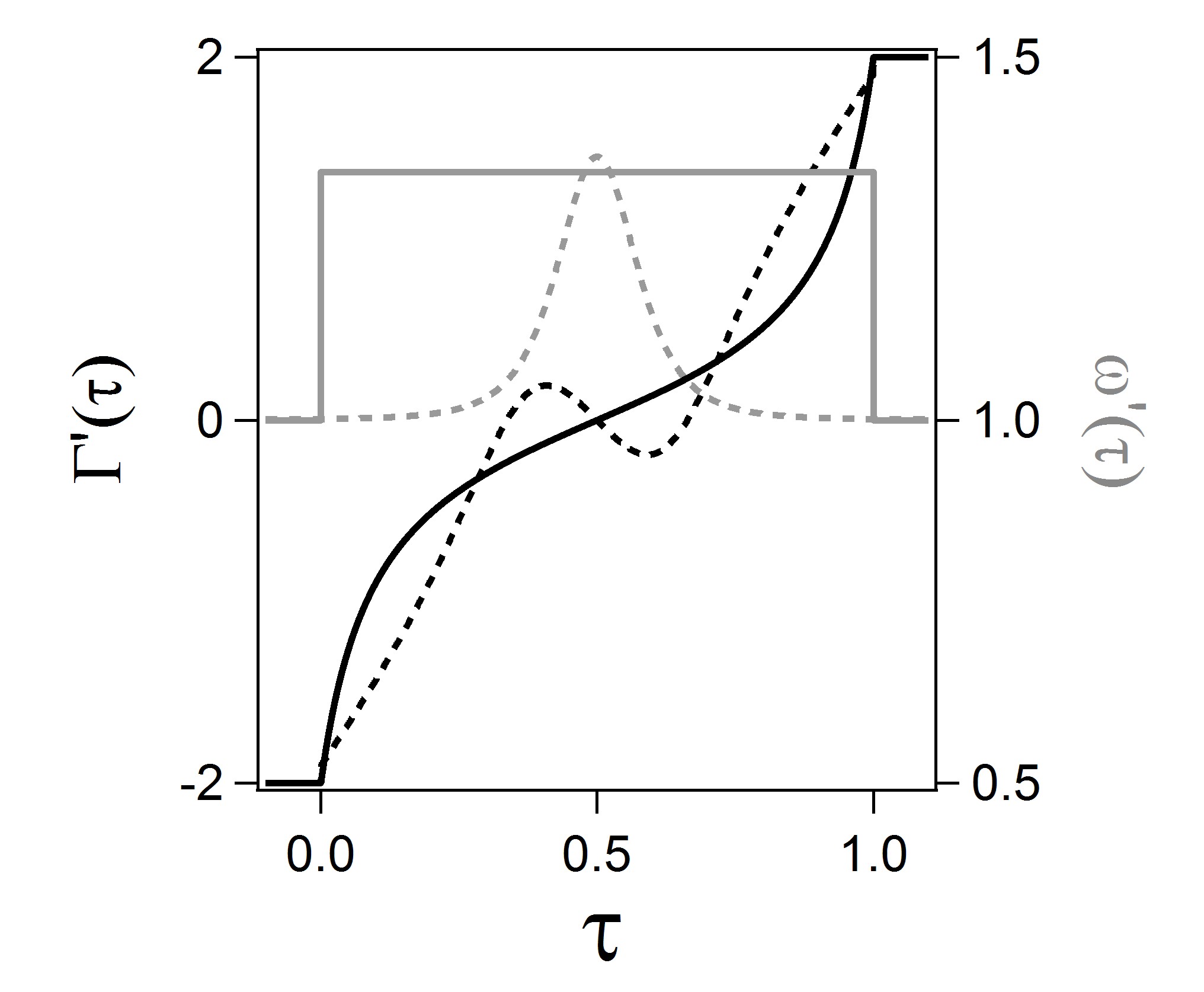}}
\caption{ Time dependence of $\Gamma^\prime(\tau)$ (black) and $\omega^\prime(\tau)$ (gray) for the superadiabatic linear (dashed line) and the superadiabatic tangent protocols (continuous line).}
\label{GammaDependencies2}
\end{figure}

\begin{figure}[t]
\centerline{\includegraphics[angle=0., width = 0.9\linewidth] {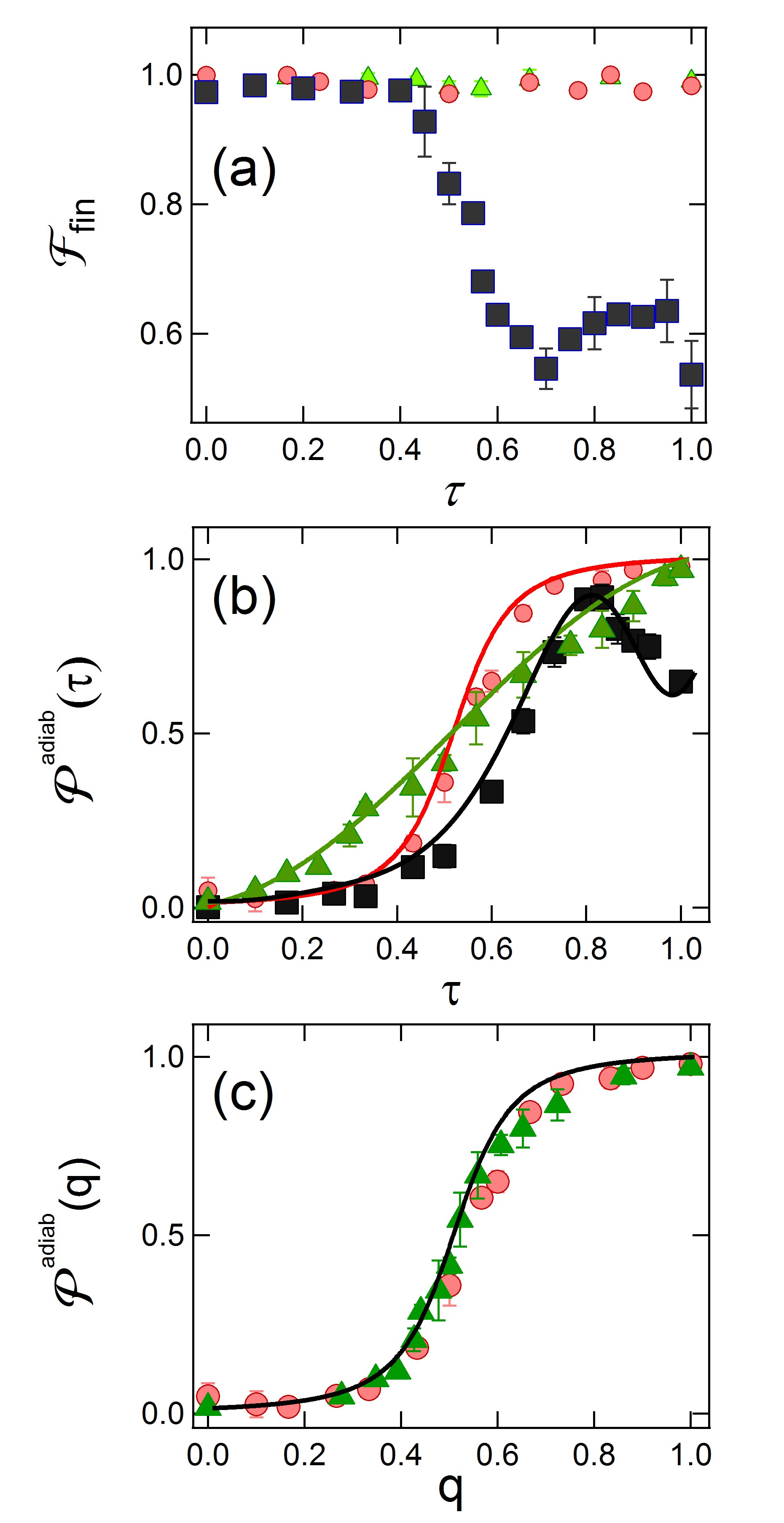}}
\caption{(a) Fidelity vs $\tau$ during the protocol for the superadiabatic linear (red circles), superadiabatic tangent (green triangles) and linear LZ (black squares) protocols in the adiabatic base.  $P^\mathrm{diab}$ as a function of the time, in (b), and of the condensate quasimomentum $q$, in (c), for the superadiabatic linear (red circles) and superadiabatic tangent (green triangles) protocols, the continuous lines derived from Eq.~\eqref{ProbDiab}. The black squares and lines show the reference LZ protocol.  Error bars are one standard deviation. Data for $\omega=0.5$  and $T=5.9$.}
\label{AdiaDia}
\end{figure}

\subsection{Diabatic and  adiabatic bases}

The dynamics of the system defined by the Hamiltonian of Eq.~\eqref{Hamiltonian} can be measured in
the diabatic and adiabatic bases. The dynamical evolution may be frozen at different times by performing a projective quantum measurement on the states of a given basis, as we explored in ref.~\cite{Zenesini2009} for our implementation of  a two-level quantum system based on  BECs in an optical lattice. We have measured the fidelity $\mathcal{F}(\tau)$ during  the time evolution of the protocol in both bases. \\
\indent For measurements in the adiabatic basis, after BEC preparation and acceleration,  the lattice depth was adiabatically reduced to zero, within 400 $\mu s$ (7.9 in natural units). The protocol fidelity $\mathcal{F}(\tau)$  was obtained as the fraction of atoms remaining in the lowest Bloch band, which  corresponds to the fraction of atoms in the $q = 0$ momentum class relative to the total atom number. Results for the linear LZ protocol and the superadiabatic protocols are shown in in Fig.~\ref{AdiaDia}(a). \\
\indent   For the measurement in the diabatic basis, after the initial loading phase and acceleration up to the time $\tau$, the atomic sample was projected onto the free-particle diabatic basis by switching off the optical lattice instantaneously (within less than 1 $\mu s$, corresponding to 0.02 in natural units). The number of atoms in the $q = 0$ and $q = 2p_{rec}$ momentum classes were measured, and the probability $\mathcal{P}^\mathrm{diab}(\tau)$ derived as the fraction of atoms in the $q = 2p_{rec}$ velocity class relative to the total atom number. Diabatic results for the linear LZ protocol and the two superadiabatic ones are shown in Fig.~\ref{AdiaDia}(b). The measured values are in good agreement with the predictions based on Eq.~\eqref{ProbDiab}.\\
\indent In Fig.~\ref{AdiaDia}(c) we show the  experimental data for $\mathcal{P}^\mathrm{diab}(\tau)$ during the evolution of $q$ at time $\tau$, and compare them to a theoretical prediction based on Eq.~\eqref{ProbDiab} and on the linear relation between $q(\tau)$ and $\Gamma(\tau)$ given by Eq.~\eqref{Gamma_Vs_q}.  Since $\omega$ is constant for the LZ and the tangent protocols, the states evolving under different superadiabatic Hamiltonians have the same parametrization in $q$.

\begin{figure}[t]
\centerline{\includegraphics[angle=0., width = 0.9\linewidth] {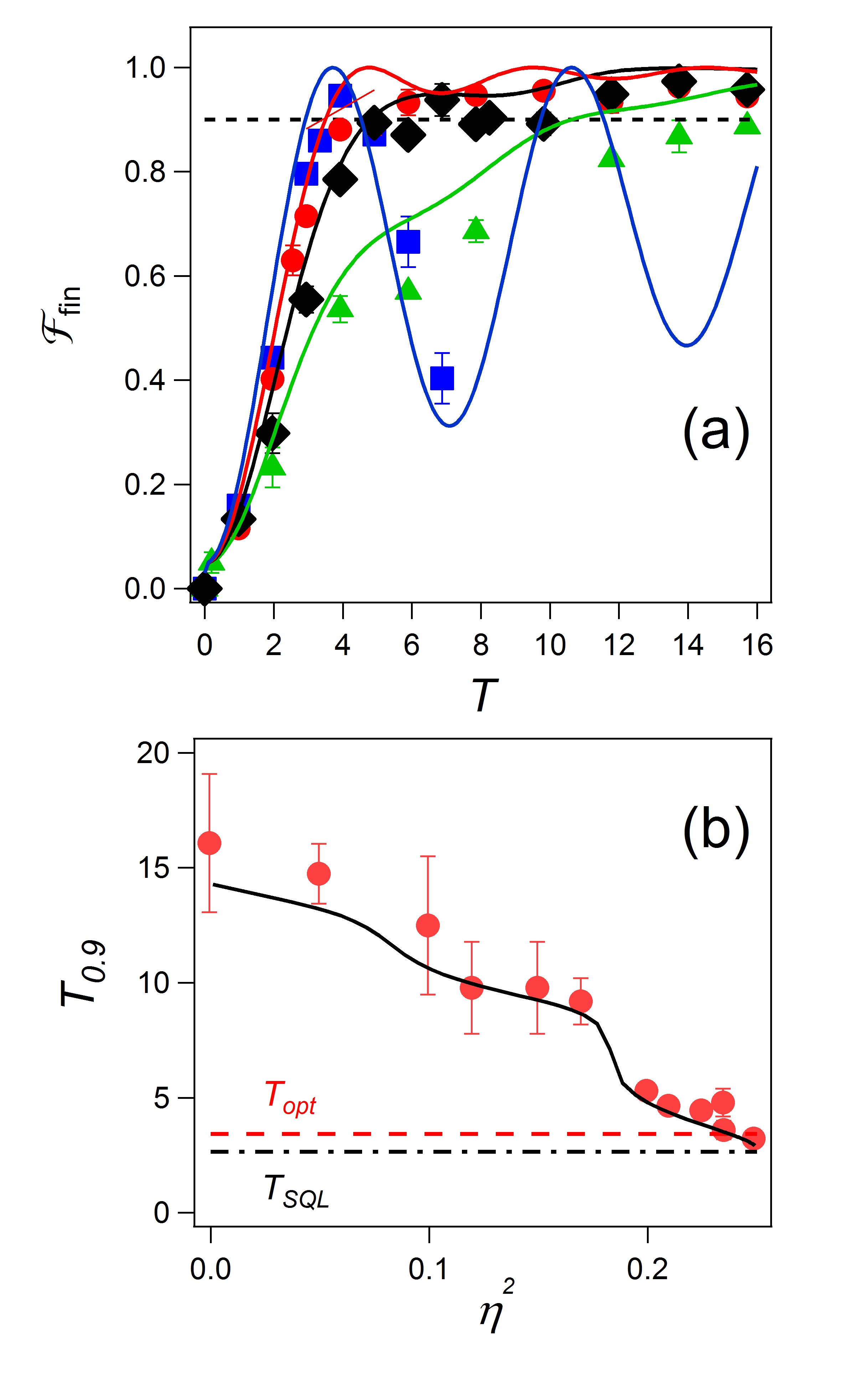}}
\caption{(Color online) (a) Measured final fidelity vs $T$ for the Roland-Cerf protocol with $\omega=0.5$ and different values of $\eta$. Green triangles $\eta^2=0.1$, black diamonds $\eta^2=0.2$, red circles $\eta= \eta_{opt}$, blue squares $\eta^2=0.249$. The continuous lines report the theoretical predictions, with the best fit obtained supposing  $\omega=0.45$. The $\mathcal{F}=0.9$  target, denoted by the dashed line,  is reached at shorter times $T$ increasing $\eta$, with an oscillating $\mathcal{F}_\mathrm{fin}(T)$ occurring for $\eta>\eta_{opt}=\sqrt{0.237954}$.   (b) $T_{0.9}$ vs $\eta^2$. The limiting value of the optimized protocol for the given $\omega$ is denoted by dashed red line, with  the $T_{SQL}$ limit  denoted by the dash-dotted black line.}
\label{RolandCerf}
\end{figure}

\section{Non-optimized protocols}
As in realistic experimental conditions parameters may be known only to within some uncertainty, protocol robustness with respect to parameter variations is relevant. The RC protocol and the superadiabatic protocols require the knowledge  of the coupling constant $\omega$ in  order to compute the parameter sweeps. We can simulate a possible experimental uncertainty by deliberately performing the protocol with parameters deviating from the optimal values.
\subsection{Roland-Cerf protocol}
The experimental implementation of the RC protocol demonstrated the high sensitivity of the final result on the protocol parameters. In order to characterize our analysis, we introduce the following parameter appearing in eq.~\ref{RolandCerfGamma}:
\begin{equation}
\eta=\epsilon\omega T_{\cal F},
\label{eta}
\end{equation}
leading to the following time dependence
\begin{equation}
\Gamma(\tau) =\frac{ 4\sqrt{1-4\eta^2}(\tau - 1/2)}{\sqrt{1 - 16\eta^2(\tau - 1/2)^2}},
\label{NORolandCerf}
\end{equation}
Notice that the above $\Gamma(\tau)$ dependence satisfies the boundary conditions $\Gamma(0)=-\Gamma(1)=2$ for any value of $\eta$. The  value $\eta_{opt}$ corresponding to the optimized Roland-Cerf protocol can be expressed as a function of $\omega$ only:
\begin{equation}
\eta_{opt}=\frac{1}{\sqrt{4 + \omega^2}}.
\label{etaopt}
\end{equation}
For $\eta \ne \eta_{opt}$, Eq.~\eqref{NORolandCerf} defines a non-optimized Roland-Cerf protocol, with $\eta^2<0.25$, as it may happen if $\omega$ is not precisely known.\\
\indent Fig.~\ref{GammaDependencies}(c) reports the $\Gamma(\tau)$ dependence for different values of  $\eta$, including the one corresponding to the optimized protocol. Notice the close resemblance between the $\Gamma(\tau)$ sweep for $\eta \approx \sqrt{0.25}$ and the composite pulse protocol of Eq.~\eqref{ggg}.\\
\indent Fig.~\ref{RolandCerf}(a)  shows the experimental results of the final fidelity ${\mathcal{F}}_\mathrm{fin}$ as a function of the duration $T$ of the protocol for different values of $\eta$. It is clear that for increasing $\eta$ the ${\mathcal{F}}_\mathrm{fin}=0.9$ threshold  is reached at decreasing values of $T$. At $\eta$ larger than $\eta_{opt}$ the threshold is reached for smaller $T$, but in that regime there are strong oscillations in the final fidelity. The measured dependence of $T_{0.9}$ on $\eta^2$ is reported in Fig~\ref{RolandCerf}(b) and compared to the results of a numerical simulation.

\begin{figure}[t]
\centerline{\includegraphics[angle=0., width = 1.1\linewidth] {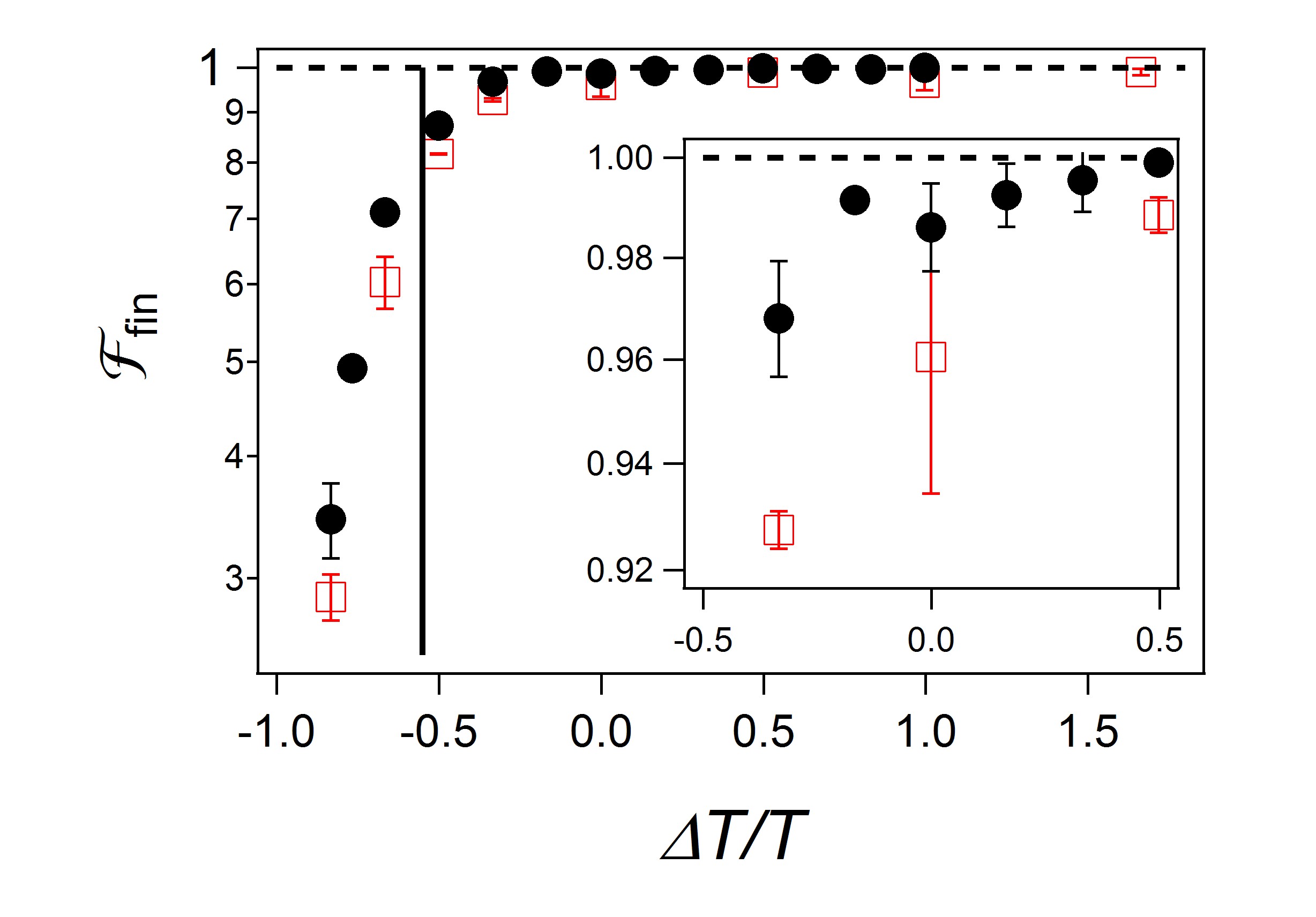}}
\caption{(Color online) Measured  fidelity $\mathcal{F}_\mathrm{fin}$ of the superadiabatic tangent protocol as a function of the relative deviation of the protocol duration from the optimum value $T$ used in Eq.~\eqref{supertan} to calculate $\omega^\prime(\tau)$ and $\Gamma^\prime(\tau)$(closed black circles), and the measured fidelity of a protocol without the superadiabatic corrections for the coupling strength, i.e. $\omega^\prime = \omega$ (open squares). The inset reports a detail of the graph on a linear scale.   The target $\mathcal{F}=1$, denoted by the horizontal dashed line,  is reached at shorter  times in the optimized protocol. The  continuous vertical  line shows the minimum time necessary to achieve $\mathcal{F}=1$ according to the quantum speed limit. Data for $\omega=0.5$ and T=5.9. }
\label{NonOptSuperAdia}
\end{figure}

\subsection{Superadiabatic protocol}
For comparison with the Roland-Cerf case we have also tested the system time evolution and ${\mathcal{F}}_\mathrm{fin}$ for the superadiabatic protocols for the case where the parameters of the superadiabatic Hamiltonian deviate from those of Eqs.~\eqref{superlin} and~\eqref{supertan}. In order to test the sensitivity of the superadiabatic tangent protocol to a variation in the control parameters, we measured the final fidelity $\mathcal{F}_\mathrm{fin}$ as a function of the deviation $\Delta T$ of the protocol duration from the duration $T$ used in Eq.~\eqref{supertan} to calculate the superadiabatic corrections. These data, shown in Fig.~\eqref{NonOptSuperAdia} as black circles, are compared to the results of a protocol without the superadiabatic corrections for the coupling strength ($\omega^\prime=\omega$), while $\Gamma^\prime(\tau)$ follows Eq.~\eqref{supertan}, including the delta-like discontinuities of  at the beginning and at the end of the protocol (red open squares data of Fig.~\eqref{NonOptSuperAdia}). This protocol, apart from the discontinuities in $\Gamma^\prime(\tau)$, is essentially a tangent protocol.
The results are summarized in Fig.~\eqref{NonOptSuperAdia}, which shows clearly that both the tangent and the superadiabatic tangent protocols are extremely robust with respect to variations of the protocol duration.  Our study demonstrates that only for large reductions of the protocol duration does the fidelity drop sharply, as otherwise the quantum speed limit would be violated. Nevertheless, the tangent protocol reaches a final fidelity $\mathcal{F}_\mathrm{fin}\simeq0.99$ for a protocol duration a factor two longer than the corresponding superadiabatic protocol. Moreover, the superadiabatic tangent protocol, ensuring a perfect following of the adiabatic ground state for $\Delta T=0$, yields $\mathcal{F}_\mathrm{fin}\gtrsim0.99$ for an increase in $T$ up to 100$\%$ of its optimum value.

\section{Best use of resources}
The protocols discussed so far, from generalized LZ sweeps to the superadiabatic protocols, can be compared in terms of various performance indicators. The first one is the capability to reach, at least in theory, a fidelity equal to 1 in a finite sweep duration. All the protocols examined here, except the standard LZ sweep, have this capability. A second indicator is the minimum time required to reach ${\mathcal{F}}_\mathrm{fin}=1$ for a given value of the coupling strength $\omega$, constant during the protocol. The composite pulse protocol realized such minimum time, $T=\pi/ \omega$ given by the quantum speed limit. Finally, a very important indicator is the robustness  of the final fidelity respect to variations of the sweep parameters. With respect to this problem, the superadiabatic protocols, based on a completely different approach, have been shown to be extremely robust. In fact, while the final fidelity for the composite pulse protocol and the RC protocol (the two protocols examined in this paper, apart from the superadiabatic ones, requiring the knowledge of the system parameters in order to compute the sweeps of $\omega$ and $\Gamma$) have a non-trivial behavior as a function of the system parameters, the superadiabatic protocols fidelity remains $\mathcal{F}_\mathrm{fin}\gtrsim0.99$ for a wide range of parameters.\\
\indent In this section the performance indicator is the minimum time required to reach a final state with ${\mathcal{F}}_\mathrm{fin}=1$, given some finite experimental resources.

\subsection{Average or peak coupling strength}
An important question in the use of quantum control protocols are the resource requirements. The superadiabatic protocols we have investigated require a well chosen  time dependence $\Gamma(\tau)$ corresponding to the  laser detuning, and in addition they require an increase of the coupling strength, from $\omega$ to $\omega^\prime(\tau)$, which corresponds to an increase of the applied laser intensity.  On the basis of the   quantum-speed limit $T_\pi \approx \pi/\omega$, that increase corresponds to a faster transfer to the target state even with a simple Rabi-pulse protocol. Therefore it is important to question whether the extra resources should be devoted to the application of the superadiabatic protocol, or to increasing the performance of a standard protocol.\\
\indent The request on the resource increase should be discussed in terms of either the peak coupling strength $\omega_{peak}$ required for the protocol realization or of the required coupling strength averaged over the applied time. We introduce the average coupling strength $\langle\omega\rangle$ defined as
\begin{equation}
\langle\omega\rangle=\int_{\tau=0}^{\tau=1}\omega^\prime(\tau)d\tau.
\end{equation}
Ref.~\cite{caneva} has pointed out that in the case of a time dependent Hamiltonian the quantum speed limit of Eq.~\eqref{qsloptimal} should depend on $\langle\omega\rangle$ instead of $\omega$.

\begin{figure}[t]
\centerline{\includegraphics[angle=0., width = 1.0\linewidth] {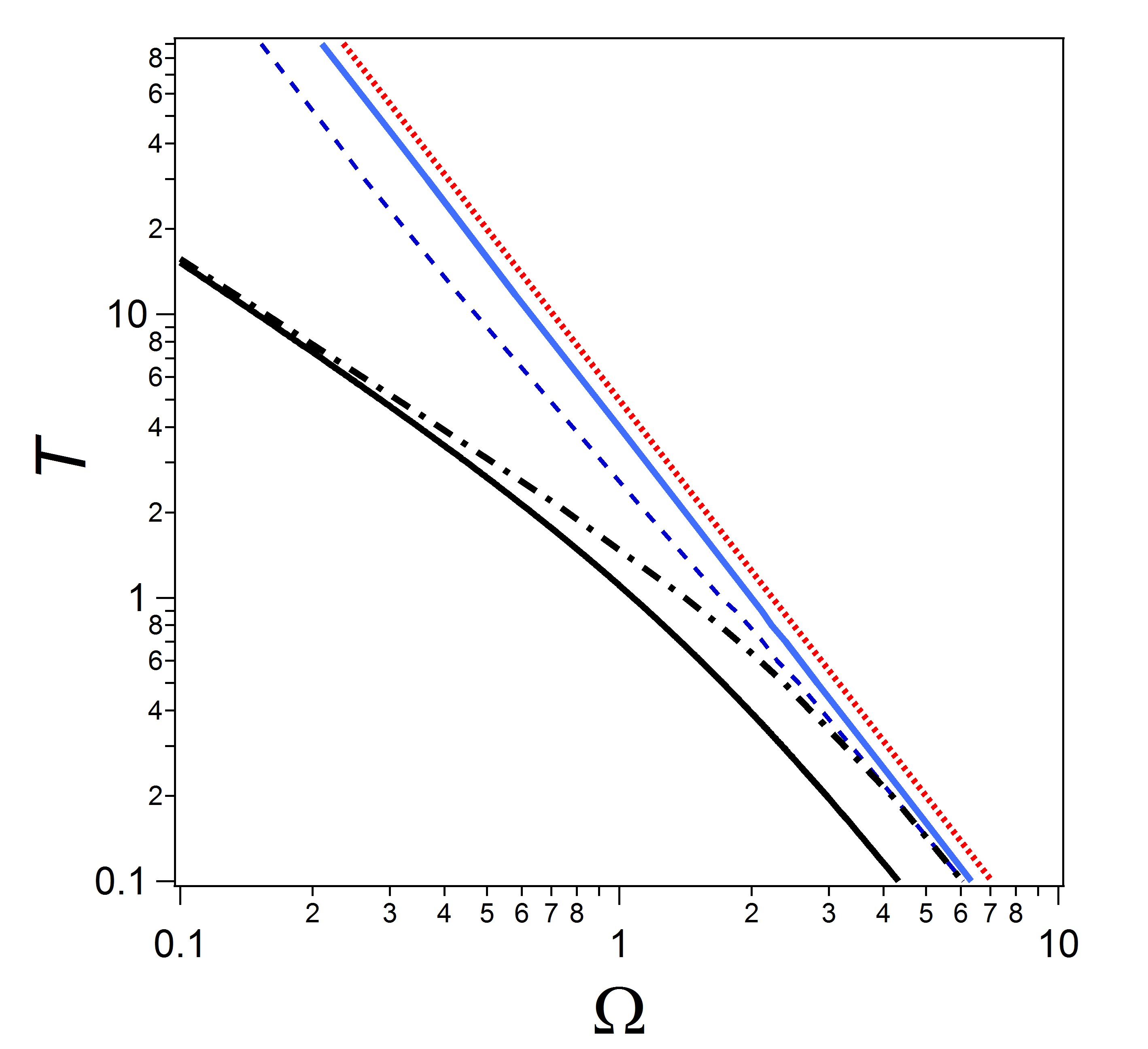}}
\caption[Fig.~1]{The red dotted line on top shows the time $T$ required for the LZ protocol to reach a 98$\%$ final fidelity vs $\Omega = \omega$, which is time-independent. For the superadiabatic linear protocol,  the blue continuous line (second from the top) shows the time $T$ required to reach 100$\%$  final fidelity   vs $\Omega = \omega_{peak}$, the minimized peak value of the time dependent $\omega^\prime(\tau)$, while for the blue dashed line (third from top)  $\Omega = \langle\omega \rangle$. For the superadiabatic tangent protocol, the black dashed-dotted line shows the time $T$ required to reach 100$\%$  final fidelity vs $\Omega= \omega^\prime$, the minimized, time-independent coupling constant.  The continuous black line represents the quantum speed limit time $T$  vs $\Omega = \omega$.}
\label{LZoptimized}
\end{figure}

\subsection{Superadiabatic linear protocol}
The  expression for $\omega^\prime(\tau)$ of Eqs.~\eqref{superlin} can be inverted in order to derive the initial $\omega$ and its peak value during the time evolution required to reach the target state within a given duration of the protocol $T$. It turns out that for each given duration there exists a unique choice of those two parameters leading to a maximum evolution speed. This analysis, applied to a range of values, is reported in Fig.~\ref{LZoptimized}. The dotted line is the prediction of the duration of the LZ protocol required to reach ${\mathcal{F}}_\mathrm{fin}=0.98$. The other lines report the results for the superadiabatic protocols, which have a final fidelity equal to 1. For a given duration $T$ of the protocol, we determine the smallest values of  initial $\omega$ and  $\langle \omega \rangle$ producing a superadiabatic linear transfer within the  time $T$. The continuous blue line reporting the  dependence of $T$ on the peak value of $\omega^\prime$ lies below that of the standard LZ protocol. At a given peak value, the superadiabatic linear protocol produces a perfect target transfer with a  35$\%$ decrease in the transfer time with respect to the LZ protocol with $\mathcal{F}_\mathrm{fin}=0.98$.\\
The  dependence of $T$ on $\langle\omega\rangle$, given by the blue dashed line, exhibits  a change in slope in the intermediate explored region, because at small values of $\omega$ the linear superadiabatic protocol imposes a  $\omega^\prime(\tau)$ which differs from $\omega$ essentially only in the region around $\tau=0.5$ ($\langle\omega\rangle \neq \omega_{peak}$). The transfer time $T$ for the superadiabatic linear protocol at a given value of $\langle\omega\rangle$ is  50$\%$ shorter than the time required by the LZ protocol with ${\mathcal{F}}_\mathrm{fin}=0.98$ for  $\langle\omega\rangle \lesssim 1$. For $\langle\omega\rangle \gg 1$ the correction of the coupling strength according to the  superadiabatic linear protocol vanishes ($\omega^\prime \approx \omega$ and $\langle\omega\rangle \approx \omega_{peak}$). \\
In the explored range of the coupling strength, the minimum time $T$ required by the superadiabatic linear protocol is larger than the quantum speed limit time (continuous black line in Fig.~\ref{LZoptimized}) by a factor between 2 ($\langle\omega\rangle \approx 5$) and 10 ($\langle\omega\rangle \approx 0.1$).

\subsection{Superadiabatic tangent protocol}
For the superadiabatic tangent protocol we also applied the optimization approach minimizing $\omega$ and $\omega^\prime$ at a given $T$ value. Note that for this protocol the coupling strength is constant, i.e. $\omega^\prime= \langle\omega\rangle$. The results for $T$ vs $\omega^\prime$ are reported in Fig.~\ref{LZoptimized} (black dashed-dotted line). Interestingly, the optimized protocol reaches the quantum speed limit for $\omega^\prime \lesssim 0.1$ and deviates at larger $\omega^\prime$. This result is quite interesting, since it means that there exists a superadiabatic protocol that ensures a perfect following of the instantaneous ground state and approaches the quantum speed limit.

\section{Conclusions}

We have explored several quantum protocols transferring the initial state of a two-level system into a final orthogonal one. The flexibility and quantum control associated with our two level system based on an optical lattice BEC configuration has allowed us to measure the final fidelity,  the minimum time required to reach the target state, and also to determine, in different bases, the temporal evolution of the wavefunction along the quantum trajectory. In addition, we have explored the robustness of the final fidelity against a variation of the Hamiltonian parameters. We have compared our experimental results to numerical simulations based on the Schr\"{o}dinger evolution of a single particle.\\
\indent Most of our protocols operate with constant $\omega$, hence the temporal dependence of $\Gamma^\prime(\tau)$ determines the overall protocol. Figures~\ref{GammaDependencies} and \ref{GammaDependencies3}, where those dependencies are depicted,  shows that the protocol efficiency is optimal when  $\Gamma^\prime(\tau)$ is nearly constant near the energy level anticrossing. Because the adiabatic theorem links the tunneling probability at the anticrossing to the time derivative of $\Gamma^\prime(\tau)$, a nearly constant value implies that the probability of leaving the adiabatic quantum state is very small, and the protocol operates with high efficiency.\\
\indent  As explored in our previous work~\cite{Bason2012}, the quantum speed limit can be approached using the superadiabatic protocol operating in the optimized regime. An important result of the present work is that the  quantum speed limit is also reached by the power law protocol with large $\alpha$. Therefore a simple modification to the LZ scheme may lead to a large improvement in operation speed. We have also discussed the important question of  the resources to be implemented for different protocols in order to reach the final state in a given time.\\
\indent While the BEC two-level configuration we used in the present investigation is not useful as a qubit for quantum computation, our results can be implemented on any qubit configuration, and in particular they could be exploited by the trap ion community where very high fidelity levels are commonly reached.

\section{Acknowledgments}
This work was supported by  the E.U. through grants No. 225187-NAMEQUAM and No. 265031-ITN-COHERENCE and by the MIUR PRIN-2009 QUANTUM GASES BEYOND EQUILIBRIUM. The authors thank J. Dalibard, P. Clad\'e, D. Gu\'{e}ry-Odelin, and M. Holthaus for fruitful discussions.

\appendix

\section{Two-level BEC Hamiltonian}
The generalized two-level LZ theory is applied to study the temporal evolution of ultracold atoms loaded into a spatially periodic potential $\frac{V_0}{2} \cos\left(2\pi x/d_L+\phi(t)\right)$. In the reference frame of the standing wave~\cite{Madison1997}, this potential becomes the sum of a periodic potential $\frac{V_0}{2} \cos\left(2\pi x/d_L\right)$ and  a slowly varying force $F$ in the presence of negligible atom-atom interactions, as is in the case of our experimental conditions~\cite{Zenesini2009}.  The characteristic energy scale of the system is the recoil energy defined as $E_{\rm rec}=\hbar\omega_{\rm rec}=\pi^{2}\hbar^{2}/2 M d_{\rm L}^{2}$. For $^{87}$Rb $\omega_{\rm rec}=2\pi\times 3.125\,\mathrm{kHz}$  defines the natural units of energy $\hbar\omega_\mathrm{rec}$ and time $1/\omega_\mathrm{rec}$ for our system.\\
\indent  The lowest energy levels of such a system are given by the quadratic dispersion relations of free particles. Momenta differing by $2 p_{rec}$ (where $p_{rec}=\hbar k_L$) are coupled through a $\omega$ coupling constant given by
\begin{equation}
\hbar \omega=\frac{V_0}{4},
\end{equation}
leading to energy bands of the quasimomentum $\tilde{q}$.
 The dynamics of ultracold atoms in a tilted optical lattice can be described by the well-known Wannier-Stark Hamiltonian~\cite{Tayebirad2010}. Under the action of the force $F$, the quasimomentum of a condensate initially prepared at  $\tilde{q}=0$  in the $n=0$ energy band scans the lower band in an oscillating motion with  the Bloch period given by $T_{Bloch}=2\pi\hbar/(Fd_L)$. At the edge of the Brillouin zone, around $\tilde{q}=\pi/(2d_L)$, where an anticrossing of order $V_0$ is present, tunneling of the condensate to the $n=1$ energy band may occur.  The full dynamics of the Wannier-Stark system can be locally approximated by a simple two-state model, with an effective Hamiltonian brought into the form of Eq.~\eqref{Hamiltonian} by subtracting the quadratic term in the momentum \cite{Tayebirad2010}. \\
\indent The time dependence of $\Gamma$ is induced through a variation of the quasimomentum $q$ of the condensate by the force $F$. The time dependence $q(t)$ leads to
\begin{equation}
\Gamma(t)=2\Gamma_0\left(\frac{q(t)}{\hbar k}-\frac{1}{2}\right),
\label{Gamma_Vs_q}
\end{equation}
where $\pm \Gamma_0$ represent the initial and final values of $\Gamma$. \\
\indent There are some limiting cases and experimental parameters  for which the simplified two-state model is not a good approximation for our system. The discrepancy  is considerable for lattice depths very large compared to  the energy scale $E_\mathrm{rec}$ of the system, for which the gap between energy bands increases leading to quasi-flat bands and localized eigenstates. Therefore, several momentum eigenstates contribute with a non-negligible amount to the lowest energy eigenstate, and one would need to take into account more components in the Hamiltonian matrix.\\
\indent While the final fidelity values are obtained performing a projective measurement after  accelerating the lattice for a full Bloch period, the $\mathcal{F}(\tau)$ fidelity is obtained by performing a projective measurement at a time $t$ smaller than $T_{Bloch}$.

\section{Experimental implementation}

In our experiments we realized an effective two-level system using ultracold atoms, forming a Bose-Einstein condensate, in an optical lattice \cite{Zenesini2009} subjected to a time-dependent force. Initially, we created Bose-Einstein condensates of $5\times10^4$ $^{87}$Rb atoms inside an optical dipole trap (mean trap frequency around 80 Hz). A one-dimensional optical lattice potential created by two counter-propagating, linearly polarized Gaussian beams was then superposed on the Bose-Einstein condensate by ramping up the power in the lattice beams to a  final $V_0$ value in the range 1 to 5 $E_{\rm rec}$. The wavelength of the lattice beams was $\lambda_L = 842$ nm, leading to a lattice constant $d_L = \lambda/2 = 421$ nm. \\\indent The experimental protocols are carried out using the techniques previously developed by us and described in detail in~\cite{Zenesini2009,Tayebirad2010}. The force is experimentally implemented by accelerating the optical lattice \cite{Kasevich}. A small frequency offset $\Delta\nu(t)$ between the two beams was introduced through the acousto-optic modulators in the setup, which allowed us to accelerate the lattice in a controlled fashion and, hence, to control the quasi-momentum $q(t)$ of the condensate. The time dependence of $\omega$ is controlled through the power of the lattice beams which determines $V_0$.\\
\indent The final state is detected by measuring the atomic momentum distributions in time-of-flight images. Our BEC has an initial width in momentum space much smaller than the width of the first Brillouin zone. This enables us to observe the full dynamics for single or multiple crossings, the only limitation being the initial momentum width of the condensates and nonlinear effects.\\
 \indent  The noise in our imaging system does not allow us to reliably to measure  fidelities $\mathcal{F} \gtrsim 0.98$. In addition,  the non-adiabaticity of the preparation and measurement protocols contributes with an infidelity on the order of 0.01.

%\bibliographystyle{naturemag}
%\bibliography{RLref}
\newpage

\end{document}